\documentclass[times]{aastex631}
\usepackage{amsmath}
\usepackage{bm}
\usepackage{ulem}

\def\dotR{\dot{\cal R}}
\def\ergs{\rm erg\,s^{-1}}
\def\kms{{\rm km\,s^{-1}}}
\def\sunm{M_\odot}
\def\sunmyr{\sunm\,yr^{-1}}
\def\bmTheta{\bm \Theta}
\def\bmJ{\bm J}
\def\bmOmega{\bm \Omega}
\def\bmm{\textbf m}
\def\bmalpha{\mathbf \alpha}
\def\bmbeta{\mathbf \beta}
\def\dotS{\dot{\cal S}}

\def\Rm{R_{-}}
\def\Rp{R_{+}}
\def\d{\mathrm{d}}
\def\Vout{V_{\rm out}}
\def\s2{\sigma_0^2}
\def\se2{\sigma_{e}^2}
\def\sunmyr{\sunm\,{\rm yr^{-1}}}

\def\tilPhi\tilde{\Phi_{0}}
\def\tilPhie\tilde{\Phi_{\rm out}}

\newcommand*{\rom}[1]{\expandafter\romannumeral}

\def\ihep{Key Laboratory for Particle Astrophysics, Institute of High Energy Physics,
Chinese Academy of Sciences, 19B Yuquan Road, Beijing 100049, China}

\def\UCASastro{School of Astronomy and Space Sciences, University of Chinese Academy of Sciences, 
19A Yuquan Road, Beijing 100049, China}

\def\UCASphy{School of Physical Sciences, University of Chinese Academy of Sciences, 
19A Yuquan Road, Beijing 100049, China}

\def\NAOC{National Astronomical Observatories, Chinese Academy of Sciences, 20A Datun Road, Beijing 100101, China}

\def\Kavli{Kavli Institute for Astronomy and Astrophysics, Peking University, Beijing 100871, China}
\def\PKUDoA{Department of Astronomy, School of Physics, Peking University, Beijing 100871, China}

\begin{document}

\title{\large Bulge Oscillation Driven by Outflows of Active Galactic Nuclei. I. Fast Outflow Case}

\author[0009-0006-2592-7699]{Yue-Chang Peng}
\affil{\ihep}
\affil{\UCASphy}

\author[0000-0001-9449-9268]{Jian-Min Wang}
\affil{\ihep}
\affil{\UCASastro}
\affil{\NAOC}
\email{wangjm@ihep.ac.cn}

\author{Yu Zhao}
\affil{\ihep}
\affil{\UCASphy}

\author{Luis C. Ho}
\affil{\Kavli}
\affil{\PKUDoA}


\begin{abstract}
There is growing evidence for star formation inside outflows of active galactic nuclei (AGNs). The formed stars are injected into bulges and give rise to perturbation of bulges. In this paper, we investigate the issues of non-rotating, spherically symmetric bulges under the perturbation of fast, massive outflows with stars formed inside. We show that the potential perturbation of outflows together with injection and dynamical friction of these stars could drive bulge oscillations. Still, we find non-zero radial velocity of bulges will be driven by the episodic outflows of AGNs and after the AGN quenched, the radial velocity will tend to zero within a timescale $\sim\tau_{\rm AGN}$, which is the AGN's lifetime. For some typical values of bulges and AGNs, we find the expansion and contraction velocities are of a few $10\,\kms$ for $10^{10}\,\sunm$ bulges and mass outflowing rate $500\,\sunm/\rm yr$, which would give observational signatures.
\end{abstract}


\section{Introduction}\label{sec:intro}
Bulges as fundamental ingredients of galaxies are one of the major investigations in astronomy. 
Because of the co-evolution between the central supermassive black holes (SMBHs) and host galaxies, bulge's kinematics provides a clue to understanding their evolution. 
As an important mechanism for co-evolution, fast, massive outflows launched from active galactic nuclei (AGN) have received widespread attention \citep{Murray1995, Proga2004, King2005}. 
There is growing evidence for star formation inside the outflows \citep{Maiolino2017, Gallagher2019}, or positive evidence of AGN feedback \citep{Xie2021, Zhuang2021, Molina2023}, thus it is obvious that the newly formed stars in outflows inevitably interact with the bulges. 
As one of the consequences of the interaction, bulge's kinematics could be changed (see  \S\ref{sec:estimation} for a simple estimation). 

Powerful outflows were observed in many contexts of AGNs
\citep{Kraemer2002, Tombesi2011, Kaastra2014, Laha2021}. 
The case of fast and massive large-scale AGN outflows, with velocities reaching $\sim 1000\, \rm \kms$ and mass outflow rates $> 1000\,\sunmyr$, extending over several kilo-parsecs, which is the typical bulge scale, has received observational supports \citep{Karouzos2016, Fluetsch2019, Lutz2020, Watts2024}. 
Theoretical studies suggest that AGN winds blowing into galaxies can shock and heat the surrounding ISM, driving slower large-scale outflows \citep{Zubovas2012, King2015, Faucher-Giguere2012}. 
Subsequently, a two-phase system is expected to form in the central few kiloparsecs as part of the ISM cools down, and the cold phase is likely to harbor star formation, with star formation rate (SFR) reaching several $100\,\sunmyr$ \citep{Zubovas2014, King2015, Cicone2014, Decataldo2024}. Numerical simulation show that these new formed stars can even coalesce to form stellar clusters with masses as large as $10^7\,\sunm$ \citep{Mercedes-Feliz2023}. 
For a more detailed discussion on outflow physics and observational characteristics, refer to the review article by \citep{Veilleux2020}.

Once these fast, massive outflows enter the bulge, their motion is mainly dominated by the bulge's gravitational potential and stellar encounters, and the latter is in the form of dynamical friction acting and capture of outflowing stars. If outflows have enough initial velocity to escape the bulge, then the gravitational potential generated by the outflows (together with outflowing stars) can be considered as an external potential perturbation to the bulge. Additionally, dynamical friction experienced by outflowing stars can also perturb the bulge. These kinematic effects brought by outflowing stars on the bulge need to be carefully explored.


In this paper, we consider the scenario where the outflowing stars can escape the bulge, and the scenario of outflows at a slower speed will be further discussed in an upcoming article. To tackle the issues of bulges perturbed by fast AGN outflows, we use the kinetic theory of stellar systems \citep[see a recent review of][]{Hamilton2024} and demonstrate how non-zero radial velocity of non-rotating spherically symmetric bulges can be excited by the injections of AGN outflows. In \S2, we provide the linear perturbation theory of bulges under AGN outflows. Numerical method and results of the bulge under perturbation are presented in \S3 and \S4, respectively. We draw conclusions in the last section.

\section{Perturbation Theory of bulges} \label{sec:theory}

\subsection{Basic considerations}\label{sec:estimation}
In order to illustrate the scientific goals of the present paper, we show two kinds of estimations of interactions between bulges and AGN outflows before detailed calculations. For an isothermal and self-gravity dominated gas sphere, 
the well-known oscillation period is $T_*\approx 1/\sqrt{\rho_*G}$, where $\rho_*$ is the mean density of stars and $G$ is the gravitational constant. 
For a typical bulge of Seyfert galaxy, 
the period of bulge's oscillation $T_{\rm bul}\approx 10^7\,M_{10}^{-1/2}R_{\rm bul,1kpc}^{3/2}\,$yr for a perturbation of gravitational potential if the fluid approximation is applied \citep{BT2008}, where $M_{10}=M_{\rm bul}/10^{10}\,\sunm$ and $R_{\rm bul,1kpc}=R_{\rm bul}/1\,{\rm kpc}$ are the mass and outer boundary of the bulge. 
This ($T_{\rm bul}$) is comparable with the episodic lifetime of AGNs $\tau_{\rm AGN}$ measured by the proximity effects of quasar absorption lines \citep[e.g.][]{Bajtlik1988} which has been shown in a relatively narrow range of $1-30\,\rm Myr$ in $z=2–3$ quasars \citep[e.g.][]{Khrykin2021}. 
This indicates that significant potential perturbations of bulges due to interaction with AGN outflows will trigger oscillations during the AGN lifetime.

Moreover, the dynamical interaction between the stars formed in AGN outflows and bulge's stars also leads to changes of bulge's kinematics as shown below.
The kinematic power of the outflows reads
\begin{equation}
L_{\rm out}=\frac{1}{2}\,{\dotR_*}V_{\rm out}^2
\simeq 3\times 10^{44}\,\dotR_{3}V_{3}^2\,\ergs,
\end{equation}
where $\dotR_{3}=\dotR_*/10^3\,\sunmyr$ is star formation rates inside the outflows, and $V_{3}=V_{\rm out}/10^3\,\kms$ is the velocities of the outflowing stars.
We assume that the star formation rates are equal to the mass rates of the outflow in this paper.
The expanding powers are given by
\begin{equation}\label{eq:luminosity}
\dot{E}_{\rm exp}\approx \frac{GM_{\rm bul}^2}{R_{\rm bul}^2}V_{\rm bul}
\simeq 9\times 10^{42}\,M_{10}^2R_{\rm bul, 1kpc}^{-2}V_{30}\,{\ergs}\approx 0.03\eta_{0.03}L_{\rm out},
\end{equation}    
where $V_{30}=V_{\rm bul}/30\,\kms$ is the expansion velocity of bulges, and 
$\eta_{0.03}=\eta/0.03$ is the conversion efficiency of the kinetic energy into bulk motion of bulges. 
The interaction efficiency depends on the outflow velocities (see Equation~(\ref{eq:a*})). 
When $V_{\rm out}$ is comparable with the dispersion velocity ($\sigma_0$) of galaxies, the interaction will be most efficient. 
In the context of fast outflows ($V_{\rm out}\gg \sigma_0$) as discussed in this paper, the efficiency is quite low.
Equation~(\ref{eq:luminosity}) suggests the necessary kinematic luminosity of AGN outflows for an observable oscillating velocity of $\sim 10\,\kms$, which is observationally testable for some nearby galaxies.
Both simple considerations show the possibility of oscillations driven by episodic outflows of AGNs through gravitational potential and dynamical interactions. 

\subsection{The Boltzmann equation}
When outflows are injected into the bulge at steady-state equilibrium, the bulge will rearrange its configurations, reaching a new equilibrium state. 
In such a case, the stellar system should be described by the Boltzmann equation with inclusion of stellar collision.
Considering that solving the time-dependent Boltzmann equation with collision terms in spherical coordinates is quite challenging, and that we only consider the case that the injection of outflowing stars are just a perturbation of bulge and give rise to radial oscillations of the bulge, we describe the bulge as a collisionless system through the Boltzmann equation, with the outflowing stars have bulk velocities significantly larger than the dispersion velocity of the bulge ($V_{\rm out}\gg\sigma_0$). 
The original stars comprising the bulge are described as the "cold" phase, whereas the outflowing stars as the "hot" phase. 
The interaction between the two phases is considered in this paper. 

We start from the Fokker-Planck equation with collisional term and source term \citep[e.g.,][]{BT2008},
\begin{equation}\label{eq:fokker-planck}
    \frac{\partial f}{\partial t} + \bm{v}\cdot\frac{\partial f}{\partial \bm x} - \nabla\Phi\cdot\frac{\partial f}{\partial \bm v} = \epsilon\Gamma(f)+\dotS,
\end{equation}
where $f$ is the distribution function of bulge stars, $(t, \bm x, \bm v)$ are the phase space coordinates, $\Phi$ is the gravitational potential, $\epsilon$ is a dimensionless parameter marking the strength of the collisional term, $\Gamma(f)$ is the encounter operator, and $\dotS$ is the source injection of outflowing stars.
We assume all stars have the same masses (of $1\,M_{\odot}$) in this paper. 
With $\epsilon \ll 1$, we take collisions as small perturbation to the system. Considering the $\epsilon$ order of quantities in the encounter operator (see Appendix~\ref{appendix:appendix.b} for details), we get
\begin{equation}\label{eq:epsilonGamma}
    \epsilon\Gamma(f) = -D_0(\Delta v_{\rm R})\frac{\partial f_0}{\partial v_{\rm R}} + \frac{1}{2}D_0\left[(\Delta v_{\rm R})^2\right] \frac{\partial^2 f_0}{\partial v_{\rm R}^2} + \frac{1}{2}D_0\left[(\Delta v_{\perp})^2\right] \frac{\partial^2 f_0}{\partial v_{\perp}^2}.
\end{equation}
where $f_0$ is the unperturbed distribution function, $v_{\rm R}$ and $v_{\perp}$ are radial and tangential velocities, respectively, $D_0(\Delta v_i)$ and $D_0[(\Delta v_i)^2]$ are the diffusion coefficients describing the expectation of the change of a unit phase space volume in $v_i$ per unit time (Appendix~\ref{appendix:appendix.b}). 
For outflows with velocity $V_{\rm out} \gg \sigma_0$, the second term is smaller than the first and third terms in the above equation. 
For a number of encounters, change of $\Delta v_{\perp}$ should be zero, indicating that only the first term (i.e. dynamical friction) takes effects along radial velocity, giving that
\begin{equation}
    \epsilon\Gamma(f) \approx - D_0(\Delta v_{\rm R})\frac{\partial f_0}{\partial v_{\rm R}}
    = - \vec{a}_*\cdot\frac{\partial f_0}{\partial \bm{v}},
\end{equation}
which is derived by Equation\,(\ref{eq:D0}). $\vec{a}_{*}$ is the acceleration of bulge stars due to collisional effect by outflowing stars. 
We have neglected the 2nd and 3rd terms on the right side of Equation\,(\ref{eq:epsilonGamma}) for the spherical cases without rotation. 
However, the 3rd term will be important if we consider rotating bulges.

Since outflowing stars' velocity distribution is completely biased to $v_{\rm R}$, it is difficult to calculate the diffusion coefficients (see Appendix~\ref{appendix:appendix.b}). 
Due to the net effect of dynamical friction is along $R$ axis, to get $\vec{a}_{*}$ (or diffusion coefficient), we only consider local encounters and use a simpler approximation by specifying the friction force of outflowing stars passing through a spherical shell equal to the driving force exerted on the spherical shell of bulge stars. 
For a Maxwell distribution of bulge stars, we take the well-known form of dynamical friction of outflowing stars as
\begin{equation}
    \vec{a}_{\rm e} = - \frac{\left(4\pi\ln\Lambda\right) G^2 \rho_{0}(M_{*}+m_{*})}{\sigma_0^2}\frac{\vec{V}_{\rm out}}{V_{\rm out}}G_x,
\end{equation}
and the Newton's third law
\begin{equation}
    \rho_{0}\vec{a}_{*} = - \rho_{\rm out}\vec{a}_{\rm e},
\end{equation}
where $\ln{\Lambda}$ is the collision parameter, $\rho_0$ is the density distribution of unperturbed bulge stars, $m_{*}$ and $M_{*}$ are bulge's single star mass and outflowing star (or star cluster) mass (see Section~\ref{sec:outflow}), $\rho_{\rm out}$ is the average density of outflowing star, $V_{\rm out}$ is the outflow velocity at $R$, and $G_x$ as a function of $V_{\rm out}$ is given in Appendix~\ref{appendix:appendix.b}.
We get the spherical shell averaged acceleration on the bulge stars
\begin{equation}\label{eq:a*}
    \vec{a}_{*} = \frac{\left(4\pi\ln\Lambda\right) G^2 \rho_{\rm out}f_{*}M_{*}}{\s2}G_x \frac{\vec{R}}{R },
\end{equation}
%
where $f_{*}=(m_{*}+M_{*})/M_{*}$. Since $M_{*}=m_{*}$ (outflowing star) or $M_{*} \gg m_{*}$ (outflowing star cluster), we simply choose $f_* \sim 1$ throughout the paper. Once we get the outflowing velocity $V_{\rm out}$, we can calculate the dynamical friction term
\begin{equation}
    \vec{a}_{*}\cdot\frac{\partial f_0}{\partial \bm{v}} = \frac{\left(4\pi\ln\Lambda\right) G^2 \rho_{\rm out}f_{*}M_{*}}{\s2}G_x v_{\rm R}\frac{\partial f_0}{\partial E},
\end{equation}
where $E$ is the energy of bulge stars. We have assumed the bulge initial distribution to be isothermal and stars in outflow mainly have radial velocity. 

Following \cite{BT2008}, we rewrite the linearized Fokker-Planck equation (Equation~(\ref{eq:fokker-planck})) into a more general form
\begin{equation}
    \frac{\partial f_1}{\partial t} + \left[ f_1,H_0 \right] + \left[ f_0,\Phi_1 \right] =  \mathcal{P}H(t).
\end{equation}
where$f_1$ is the perturbation of distribution function (DF),  $\left[ \cdot,\cdot \right]$ is the Poisson bracket, $H_0$ is the Hamiltonian of the unperturbed system, $\Phi_1$ is the perturbing potential containing both external and response potential, we introduce $\mathcal{P}$ as the total function denoting source and dynamical friction terms due to the outflow interaction
\begin{equation}
    \mathcal{P} = \dot{f}_{\rm out} - \vec{a}_{*}\cdot\frac{\partial f}{\partial \bm{v}},
\end{equation}
where the first term ($\dot{f}_{\rm out}$) is the injection rates of outflowing stars, and the second term is the change rates of the distribution function due to dynamical friction between the formed stars inside AGN outflows and bulge stars. $H(t)$ is the step function defined as
\begin{equation}
    H(t) = 
    \begin{cases}
    1, & \text{if } 0 < t \le \tau_{\rm AGN} \\
    0, & \text{else} \\
    \end{cases}
\end{equation}
and $\tau_{\rm AGN}$ is the AGN lifetime with outflow. 
Because the typical outflow speed ($\sim 1000\,\rm \kms$) results in a time ($\tau_{\rm cross}\sim 10^6\, \rm yr$) much less than typical $\tau_{\rm AGN}$ for outflow to across the bulge and reach the outer boundary, we simply use a step function to represent the effect of the outflow action. Though the derivation of $\mathcal{P}$ is obvious, there's little exploration by previous studies. 

\subsection{Response Kernel}
In this section, we use the time-dependent method described by \cite{Murali1999} and \cite{Dootson2022} to get the time-dependent solution of the Boltzmann equation.
Using Jeans theorem, the unperturbed $f_0$ only depends on energy $E$ and angular momentum $L$, thus we can simplify linearized Boltzmann equation using the angle-momentum coordinates ($\mathbf{\Theta}, \mathbf{J}$) as (see Appendix~\ref{appendix:appendix.a})
\begin{equation}
    \frac{\partial f_1}{\partial t} + \mathbf{\Omega}\cdot\frac{\partial f_1}{\partial \mathbf{\Theta}} - \frac{\partial f_0}{\partial \mathbf{J}}\cdot\frac{\partial \Phi_1}{\partial \mathbf{\Theta}} = \mathcal{P}H(t),
\end{equation}
where $\bm\Omega=\partial H_0/\partial \bmJ$. We expand $f_1, \Phi_1, \mathcal{P}$ as
\begin{gather}
    f_1 = \sum_{\bmm}f_{\bmm}(\bmJ,t)e^{i\bmm\cdot\bmTheta}, \\
    \Phi_1 = \sum_{\bmm}\Phi_{\bmm}(\bmJ,t)e^{i\bmm\cdot\bmTheta}, \\
    \mathcal{P} = \sum_{\bmm}\mathcal{P}_{\bmm}\left(\bmJ\right)e^{i\bmm\cdot\bm\Theta},
\end{gather}
where $\bmm=n,k,m$ is the angle index. For a given index $\bmm$, the above Boltzmann equation converts to
\begin{equation}
    \frac{\partial f_{\bmm}}{\partial t} + i\bmm\cdot \left(\bmOmega f_{\bmm} - \frac{\partial f_0}{\partial \bmJ}\Phi_{\bmm} \right) = \mathcal{P}_{\bmm}H(t),
\end{equation}
and the solution is
\begin{equation}\label{eq:f_n}
    f_{\bmm}(\bmJ,t) = \int_{-\infty}^{t}\d t'\times \left[\mathcal{P}_{\bmm}H(t') + i\bmm\cdot\frac{\partial f_0}{\partial \bmJ}\Phi_{\bmm}\left(\bmJ,t'\right) \right]e^{-i\bmm\cdot\bmOmega(t-t')}.
\end{equation}

Because in our scenario, outflow materials can escape the bulge, 
we separate the potential perturbation into external perturbation $\Phi_{\rm e}$ (i.e. potential due to materials that can escape the bulge) and response of the self-gravity of bulge stars $\Phi_{\rm s1}$
\begin{equation}
    \Phi_{1} = \Phi_{\rm e} + \Phi_{\rm s1}.
\end{equation}
Suppose $(\Phi_{\bmalpha},\rho_{\bmalpha})$ is a potential-density basis pair \citep[e.g.,][]{BT2008} with a generalized orthogonal relation
\begin{equation}
    \Delta_{\bmalpha\bmbeta} = -\int \d^3 \mathbf{x} \Phi_{\bmalpha}^{*}\rho_{\bmbeta},
\end{equation}
with $\Phi_{\bmalpha}^{*}$ denote the conjugation, $\Delta_{\bmalpha\bmbeta}$ is the orthogonal matrix and $(\Phi_{\bmalpha},\rho_{\bmalpha})$ satisfy the Poisson equation.
We can expand the external perturbation and self-gravity perturbation using the basis
\begin{gather}
    \Phi_{\rm e} = \sum_{\bmalpha}A_{\bmalpha}(t) \Phi_{\bmalpha}, \\
    \Phi_{\rm s1} = \sum_{\bmalpha}B_{\bmalpha}(t) \Phi_{\bmalpha}, \\ 
    \Phi_{1} = \sum_{\bmalpha}\left[A_{\bmalpha}(t)+B_{\bmalpha}(t)\right]\Phi_{\bmalpha}.
\end{gather}
Generally speaking, the basis is a function of spherical coordinates $(R,\theta,\phi)$
\begin{equation}
    \Phi_{\bmalpha} = \Phi_{p}(R)Y_{lm}(\theta,\phi),
\end{equation}
with $\bmalpha=(p,l,m)$, and $Y_{lm}(\theta,\phi)$ is the spherical harmonics. Since we only consider spherical symmetric case, only the zero order of spherical harmonics remains, indicating $l=m=0$ \citep[see][]{Palmer1994}. Therefore, we indiscriminately use $\Phi_{\bmalpha}$ to represent $\Phi_{p}$ in the following text.

We multiply Equation~(\ref{eq:f_n}) with $e^{i\bmm\cdot\bmTheta}\Phi_{\bmbeta}^{*}$ and sum over all index $\bmm$ then integrating over space and velocity
\begin{equation}\label{eq:matrix-equation}
    \Delta_{\bmbeta\bmalpha}B_{\bmalpha}(t) = \\ \int_{-\infty}^{t}\d t' \mathcal{K}_{\bmbeta\bmalpha}(t-t')\left[A_{\bmalpha}(t')+B_{\bmalpha}(t')\right] + \int_{-\infty}^{t} \d t' \mathcal{D}_{\bmbeta}(t-t')H(t'),
\end{equation}
where
\begin{equation}\label{eq:response-matrix}
    \mathcal{K}_{\bmbeta\bmalpha}(t-t') = -i(2\pi)^3 \int \d^3\bmJ\sum_{\bmm}\bmm\cdot\left(\frac{\partial f_0}{\partial \bmJ}\right)e^{-i\bmm\cdot\bmOmega(t-t')}\Phi_{\bmbeta,\bmm}^{*}\Phi_{\bmalpha,\bmm},
\end{equation}
is the response kernel function, and
\begin{gather}
    \Phi_{\bmalpha,\bmm} = \frac{1}{(2\pi)^3}\int \Phi_{\bmalpha}e^{-i\bmm\cdot\bmTheta}\d^3\bmTheta, \\
    \mathcal{D}_{\bmbeta}(t-t') = - (2\pi)^3\int \d^3\bmJ \sum_{\bmm} \mathcal{P}_{\bmm}(\bmJ)e^{-i\bmm\cdot\bmOmega(t-t')}\Phi_{\bmbeta,\bmm}^{*}.
\end{gather}
we can calculate the time evolution of bulge's response to perturbation once $f_{0}, \dot{f}_{\rm out}$ is known.

\subsection{AGN Outflows}\label{sec:outflow}

There are lots of theoretical models for bulge's density distribution, such as Plummer and two-power density models \citep{BT2008}. For simplicity, we consider the unperturbed distribution function of bulge stars to be isothermal, with the corresponding density $\rho_0$ and potential $\Phi_0$ as \citep{BT2008}
\begin{gather}
    f_0 
    = \frac{1}{(2\pi)^{\frac{5}{2}} GR_{\rm bul}^2\sigma_0}\exp\left(-\frac{E}{\s2}\right), \\
    \rho_{0} = \frac{\s2}{2\pi GR^2}, \\
    \Phi_{0} = 2\s2\left[\ln{\left(\frac{R}{R_{\rm bul}}\right)} - 1 \right],
\end{gather}
where $\sigma_0$ is the velocity dispersion of bulge and $R_{\rm bul}$ is the outer boundary of bulge. 
We should note that in our model, $R$ ranging from bulge's inner boundary $R_{\rm in}$ to $R_{\rm bul}$, which will lead to some additional terms in potential distribution of isothermal sphere when directly solving Poisson's equation. We show that this issue could be fixed in Appendix~\ref{appendix:appendix.c}

Though AGN outflows may play positive or negative roles in star formation in bulges, we only consider star formation in cold gas of AGN outflows \citep{Zubovas2014}. These stars can either be unrelated to each other or form clusters due to originating from the same cold gas cloud with $M_{\rm cluster}\sim 10^6\,M_{\odot}$ \citep[such as super star clusters, see][]{Ho1997, Keto2005}. The formed stars follow the velocity of the outflows, and have probability of mixing into bulge stars once captured by the bulge's potential. The time evolution of radial velocity of certain outflowing stars is dominated by gravitational potential and dynamical friction
\begin{gather}\label{eq:vout}
    \frac{\d \Vout}{\d t} = - \frac{2\s2}{R} - a_{e}, \\
    \frac{\d R}{\d t} = \Vout. 
\end{gather}
We numerically solve the above equations to get the radial distribution of outflowing velocity $\Vout$. If dynamical friction is small compared to gravitational force, we can neglect $a_{e}$ and solve above equations analytically
\begin{equation}\label{eq:analys_Vout}
    \Vout = \sqrt{V_{\rm out,0}^2 - 4\s2\ln{\left(\frac{R}{R_{\rm in}}\right)}},
\end{equation}
where $V_{\rm out,0}$ is the outflowing velocity at bulge's inner boundary $R_{\rm in}$. We can see there's a maximum distance where the outflow velocity reaches zero. For the parameters used in this paper, the minimum $V_{\rm out,0}$ for the outflow to reach $R_{\rm bul}$ is slightly less than $900 \, \rm km\,s^{-1}$.
We show the numerically solved outflow velocity in the left panel of Figure~\ref{fig:vout&basis} with different $V_{\rm out,0}$. 

The mass injection rate of outflowing stars
\begin{equation}\label{eq:outflow_conserv}
    \dot{M}_{\rm out} = 4\pi R^2 V_{\rm out}\rho_{\rm out}, 
\end{equation}
should be conserved, giving the relation between the outflowing stellar density $\rho_{\rm out}$ and the outflowing velocity $V_{\rm out}$.
We should note that the actual outflow in the bulge may not be stars, but also the star-forming cold gas \citep{Zubovas2014}. 
However, we simplify the above problem by representing the star formation rate $\dotR_*$ as an equivalent star injection rate $\dot{M}_{\rm out}$, this means all the mass of cold gas is converted to outflowing stars. Once $\Vout$ is known, we can get the density perturbation distribution $\rho_{\rm out}$ by Equation~(\ref{eq:outflow_conserv}) and the corresponding potential perturbation $\Phi_{\rm e,out}$ by Poisson's equation.

Meanwhile, since the mass of the outflow mainly comes from the gas in the bulge, the decrease in gas composition within the bulge will have additional impact on the bulge's original gravitational potential. 
We assume a simple model that the potential due to mass-loss during AGN lifetime follows its original distribution
\begin{equation}
    \Phi_{\rm e,bul}(t) 
    =- \frac{G\dot{M}_{\rm out}}{R_{\rm bul}}\left[\ln{\left(\frac{R}{R_{\rm bul}}\right)}-1 \right]\times
    \begin{cases}
    t, & \text{if } 0 < t \le \tau_{\rm AGN} \\
    \tau_{\rm AGN}, & \text{else} \\
    \end{cases}
\end{equation}
and the total mass lost within $\tau_{\rm AGN}$ equals to the mass of the gas inside the bulge
\begin{equation}\label{agn_life}
    \dot{M}_{\rm out}\tau_{\rm AGN} = f_{\rm g}M_{\rm bul},
\end{equation}
where $f_{\rm g}$ is the gas fraction of the bulge's mass. In the subsequent calculations, we fix $f_{\rm g}=0.1$, corresponding to a gas mass of approximately $2\times10^9\,M_\odot$, while $\tau_{\rm AGN}$ varies with $\dot{M}_{\rm out}$. Thus, the loss of gas will not have a significant impact on the bulge.

Therefore, the total external potential $\Phi_{\rm e}$ should come from the sum of these two contributors: the gas lost from the bulge $\Phi_{\rm e,bul}$ and the steady outflowing distribution $\Phi_{\rm e,out}$
\begin{equation}
    \Phi_{\rm e}(t) = \Phi_{\rm e,out}H(t) + \Phi_{\rm e,bul}(t).
\end{equation}
where again $H(t)$ is the step function.

We set the injection distribution function $\dot{f}_{\rm out}$ in the Boltzmann equation as
\begin{equation}\label{eq:injection_func}
    \dot{f}_{\rm out} = \frac{\dot{M}_{\rm out}}{(2\pi)^{3/2} \sigma_{\rm e}^3} \frac{\rho_{\rm out}}{\int 4\pi R^2\rho_{\rm out}\d R }\exp\left[-\frac{(\bm{v}-\vec{V}_{\rm out})^2}{2\sigma_{\rm e}^2}\right].
\end{equation}
where $\sigma_{\rm e}$ is the velocity dispersion of outflowing stars. We have made the total injection rate $\dot{M}_{\rm out}$ conserve.

\subsection{Bulge Radial Velocity}

Once we get $f_1$, to calculate the radial velocity by perturbation, we need to integrate $f_1$ in velocity space. We use the definition of polar coordinates $(v_{\rm R}, v_{\theta}, v_{\phi})$ for velocity space \citep{BT2008}
\begin{gather}
    v_{\rm R} = v\cos\eta, \\
    v_{\theta} = v\sin\eta\cos\psi, \\
    v_{\phi} = v\sin\eta\sin\psi,
\end{gather}
where $v$ is the absolute value of star's velocity, $\eta$ is the angle between star's radial vector and velocity vector, $\psi$ is the angle between the normal vector of star's orbital plane and the equator of spherical coordinate. Meanwhile, using the relation between velocity and energy $E$ and angular momentum $L$
\begin{gather}
    E = \Phi(R) + \frac{1}{2}v^2, \\
    L^2 = R^2v^2\sin^2\eta = \frac{1}{2}R^2v^2-\frac{1}{2}R^2v^2\cos2\eta,
\end{gather}
we can calculate the momentum $\overline{\rho_{\rm s1} v_{\rm R}}$ by converting integral over velocity to $E,L^2$ as 
\begin{align}\label{eq:v_r}
    \overline{\rho_{\rm s1} v_{\rm R}} & = \int f_1(\bmJ,\bmTheta,t)v_{\rm R}\d v_{\rm R}\d v_{\theta}\d v_{\phi}, \\
    & = 2\pi \int f_1(\bmJ,\bmTheta,t) v\cos\eta v^2\sin\eta \d v\d\eta, \\ 
    & = \frac{\pi}{R^2}\int f_1(\bmJ,\bmTheta,t) \d E\d L^2.
\end{align}
The results are consistent with those obtained by integration using the given orbit's $v_{\rm R}$ and $v_{\perp}$ (see Equation~(\ref{eq:orbital_vr_vt})). Here we have used the relation
\begin{equation}
    \d v_{\rm R}\d v_{\theta}\d v_{\phi} = \frac{\partial(v_{\rm R},v_{\theta},v_{\phi})}{\partial (v,\eta,\psi)} \d v\d\eta\d\psi = 2\pi v^2\sin\eta \d v\d\eta.
\end{equation}
To get $f_1$, we use the summation form
\begin{equation}
    f_1(\bmJ,\bmTheta,t) = \sum_{\bmm}f_{\bmm}(\bmJ,t)e^{i\bmm\cdot\bmTheta} = \sum_{n}f_{n}(\bmJ,t)e^{i(n\theta_{\rm R})}, 
\end{equation}
where we have used $\bmm=(n,0,0)$ for spherical symmetry perturbation, thus only the $\theta_{\rm R}$ component of $\bmTheta$ remains. And $f_{n}(\bmJ,t)$ can be calculated by Equation~(\ref{eq:f_n}), thus response density and velocity momentum can be expressed as
\begin{gather}\label{eq:rhos1-vr}
    \rho_{\rm s1} = \frac{\pi}{R^2}\int \frac{\d E\d L^2}{v_{\rm R}} \sum_{n}f_{n}(\bmJ,t)e^{i(n\theta_{\rm R})}, \\
    \overline{\rho_{\rm s1} v_{\rm R}} = \frac{\pi}{R^2}\int \d E\d L^2 \sum_{n}f_{n}(\bmJ,t)e^{i(n\theta_{\rm R})}.
\end{gather}

\section{Numerical Method}\label{sec:numerical_method}

\begin{figure}
    \centering
    \includegraphics[width=0.98\textwidth]{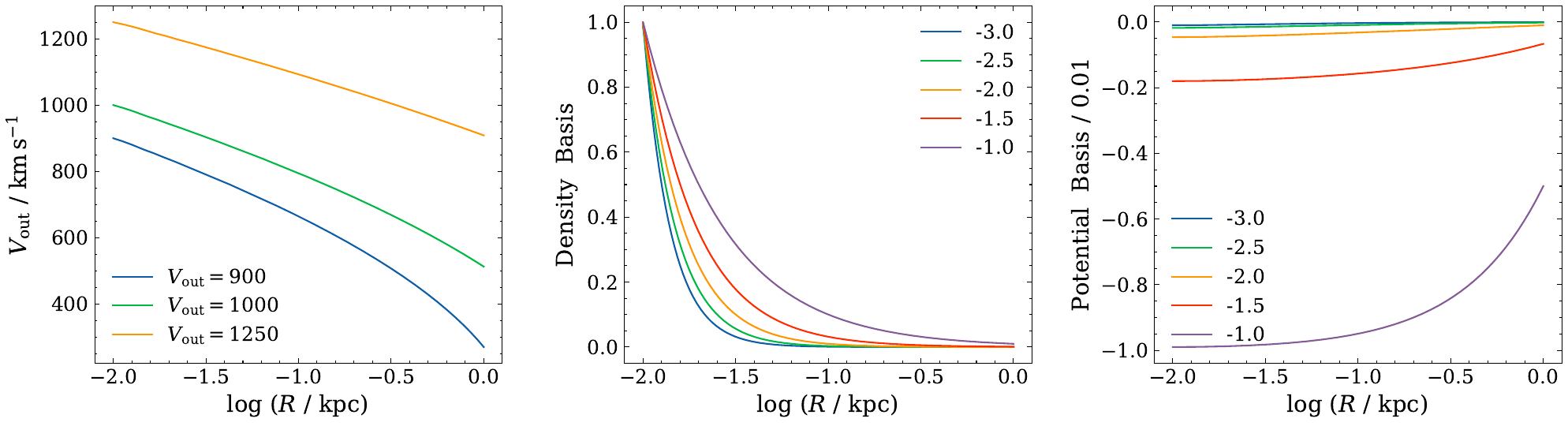}
    \caption{Left: The distribution of outflow velocity with $R$ obtained from Equation~(\ref{eq:vout}), corresponding to different outflow velocities at inner boundary. Middle: The density basis functions with different power-law indexes ranging between $\left[-3,-1\right]$, and the values at the inner boundary are normalized to $1$. Right: The potential basis functions are obtained from the density basis functions by solving Poisson's equation analytically. }
    \label{fig:vout&basis}
\end{figure}

Once $\Phi_{\rm e}$ is known, with a given number of ${\bmalpha}$ and basis function, we can get $A_{\bmalpha}$ by optimization approach. Given the number of parameters, the performance may vary with different bases. However, we leave this as a detail for further investigation and do not discuss it in this article. For simplicity, we use power-law functions with different power-law indices as the density basis, and solve Poisson's equation to get the corresponding potential basis. We present the density and potential basis, as well as the radial distribution of different outflow velocities at $R_{\rm in}$ in Figure~\ref{fig:vout&basis}.
The expansion coefficients of $\mathcal{P}$ and $\Phi_{\bmalpha}$ with respect to $e^{i(n\theta_{\rm R})}$ are
\begin{gather}\label{eq:expansion}
        \Phi_{\bmalpha,\bmm} = \Phi_{\bmalpha,n} = \frac{1}{2\pi}\int_{0}^{2\pi} \Phi_{\bmalpha}(R) e^{-i(n\theta_{\rm R})} \d \theta_{\rm R},\\
        \mathcal{P}_{\bmm} = \mathcal{P}_{n} = \frac{1}{2\pi}\int_{0}^{2\pi} \mathcal{P}(R) e^{-i(n\theta_{\rm R})}\d\theta_{\rm R}.
\end{gather}
where $\theta_{\rm R}$ is (see Appendix~\ref{appendix:appendix.a})
\begin{equation}
    \theta_{\rm R} 
    = \int_{\mathcal{L}} \frac{\Omega_{\rm R}\d R}{\sqrt{2(E-\Phi)-L^2/R^2}},
\end{equation}
and $\Omega_{\rm R}$
\begin{equation}
    \frac{\pi}{\Omega_{\rm R}} = \int_{R_{-}}^{R_{+}}\frac{\d R}{\sqrt{2(E-\Phi)-L^2/R^2}},
\end{equation}
where $\Omega_{\rm R}$ is the radial frequency of a given star, the above integral is performed along the orbit $\mathcal{L}$ starting from the pericenter, and moving to the apocenter $\theta_{\rm R}=\pi$, covering one full orbit with $\theta_{\rm R}$ changing from $0$ to $2\pi$. Given the energy $E$ and the square of the angular momentum $L^2$ of the star on the orbit, the shape of the orbit can be determined, and in this case, $\theta_{\rm R}$ is only dependent on $R$. We discretize $\theta_{\rm R}$ and express Equation~(\ref{eq:expansion}) as
\begin{equation}
    \Phi_{\bmalpha,n}  = \frac{1}{2\pi}\int_{0}^{2\pi} \Phi_{\bmalpha}\left[R(\theta_{\rm R})\right] e^{-i(n\theta_{\rm R})} \d \theta_{\rm R}
    = \frac{1}{N}\sum_{k=1}^{N}\Phi_{\bmalpha}\left[R\left(\frac{2\pi k}{N}\right)\right]e^{-i\frac{2\pi k n}{N}}.
\end{equation}
and
\begin{align}
    \mathcal{P}_{n} & = \frac{1}{N}\sum_{k}\mathcal{P}\left[R\left(\frac{2\pi k}{N}\right)\right]e^{-i\frac{2\pi k n}{N}}.
\end{align}
where $k, n$ are integers. It can be seen that if the total grid number $N$ equals to the number of selected $n$, the above equation represents the discrete Fourier transform of $\Phi_{\bmalpha}$. Therefore, given $\bmalpha, E, L^2$, we can calculate the expansion coefficients using the discrete Fourier transform, and those expansion coefficients are defined on grids of $\theta_{\rm R}$ ranging from $0-2\pi$, corresponding to $R$ when giving the star's orbit.

To calculate the response kernel function (Equation~(\ref{eq:response-matrix})), we convert the integral over $\bmJ$ to $(E,L^2)$
\begin{equation}
    \d^3 \bmJ = \frac{\partial (J_{\rm R},J_{\theta},J_{\phi})}{\partial (E,L,\eta)}\d E\d L\d\eta = \frac{\d E\d L^2}{\Omega_{\rm R}}.
\end{equation}
where $(J_{\rm R},J_{\theta},J_{\phi})$ are actions for spherical symmetric systems \citep[for more details, see][]{Palmer1994, BT2008},
and then convert the integral to $(R_{-},R_{+})$ similar to \citep{Dootson2022}, where $(R_{-},R_{+})$ are pericenter and apocenter of certain orbit, respectively. The energy and angular momentum are
\begin{equation}
    E = \frac{R_{+}^2\Phi_{+}-R_{-}^2\Phi_{-}}{R_{+}^2-R_{-}^2}, 
    \quad 
    L = \sqrt{\frac{2\left(\Phi_{+}-\Phi_{-}\right)}{R_{-}^{-2}-R_{+}^{-2}}},
\end{equation}
where $\Phi_{\pm}=\Phi(R_{\pm})$ is the unperturbed potential at $R_{\pm}$. 
The kernel matrix in Equation~(\ref{eq:response-matrix}) can be written as
\begin{align}
    \mathcal{K}_{\bmbeta\bmalpha}(t-t') 
    & = -i(2\pi)^3 \int \d^3\bmJ \sum_{\bmm}\bmm\cdot\left(\frac{\partial f_0}{\partial \bmJ}\right)e^{-i[\bmm\cdot\bmOmega(t-t')]}\Phi_{\bmbeta,\bmm}^{*}\Phi_{\bmalpha,\bmm}, \\ 
    & = -i(2\pi)^3 \int \d E\d L^2 \sum_{n}n\frac{\partial f_0}{\partial E}e^{-i[n\Omega_{\rm R}(t-t')]}\Phi_{\bmbeta,\bmm}^{*}\Phi_{\bmalpha,\bmm}, \\ 
    & = -i(2\pi)^3 \int \frac{\partial (E,L^2)}{\partial (R_{-},R_{+})}\d R_{-}\d R_{+} \sum_{n} n\frac{\partial f_0}{\partial E}e^{-i[n\Omega_{\rm R}(t-t')]}\Phi_{\bmbeta,\bmm}^{*}\Phi_{\bmalpha,\bmm}.
\end{align}
and 
\begin{align}
    \mathcal{D}_{\bmbeta}(t-t') 
    & = - (2\pi)^3\int \d^3\bmJ \sum_{\bmm} \mathcal{P}_{\bmm}(\bmJ)e^{-i[\bmm\cdot\bmOmega(t-t')]}\Phi_{\bmbeta,\bmm}^{*}, \\ 
    & = - (2\pi)^3\int \frac{\partial (E,L^2)}{\partial (R_{-},R_{+})}\d R_{-}\d R_{+}\sum_{\bmm} \mathcal{P}_{\bmm}(\bmJ)e^{-i[\bmm\cdot\bmOmega(t-t')]}\frac{\Phi_{\bmbeta,\bmm}^{*}}{\Omega_{\rm R}}.
\end{align}
with the Jacobian matrix calculated by
\begin{align}
    \frac{\partial E}{\partial R_{+}} & = \frac{1}{\left(R_{+}^2-R_{-}^2\right)^2}\left[\left(2R_{+}\Phi_{+}+R_{+}^2\frac{\partial \Phi_{+}}{\partial R}\right)\left(R_{+}^2-R_{-}^2\right)-2R_{+}\left(R_{+}^2\Phi_{+}-R_{+}^2\Phi_{-}^2\right) \right], \\
    \frac{\partial L^2}{\partial R_{+}} & = \frac{1}{\left(R_{+}^2-R_{-}^2\right)^2}\left[-4R_{+}R_{-}^4\left(\Phi_{+}-\Phi_{-}\right)+2R_{+}^2R_{-}^2\frac{\partial \Phi_{+}}{\partial R_{+}}\left(R_{+}^2-R_{-}^2\right) \right],
\end{align}
and exchange $(+,-)$ to get $\partial E/\partial R_{-}, \partial L^2/\partial R_{-}$ respectively. 
Integrating over all $(R_{+},R_{-})$ we can calculate the perturbation response at a given time. We use the similar method to calculate other quantities.
By numerically solving Equation~(\ref{eq:matrix-equation})
\begin{multline}\label{eq:matrix-equation-grid}
    \left[\Delta_{\bmalpha\bmbeta}-\Delta\tau\mathcal{K}_{\bmalpha\bmbeta}(0) \right]B_{\bmbeta}(\tau_{n}) = \Delta\tau\mathcal{K}_{\bmalpha\bmbeta}(0)A_{\bmbeta}(\tau_n) + \Delta\tau\sum_{k=0}^{n-1}\mathcal{K}_{\bmalpha\bmbeta}(\tau_n-\tau_k)\left[A_{\bmbeta}(\tau_k)+ B_{\bmbeta}(\tau_k)\right] + \\ \Delta\tau\sum_{k=0}^{n}\mathcal{D}_{\bmbeta}(\tau_n-\tau_k)H(\tau_k),
\end{multline}
where $\Delta\tau$ is the temporal grid length, $\tau_n$ denotes the temporal grids.
we can get the time evolution of bulge response to outflow perturbation.

To get the bulge's radial velocity (Equation~(\ref{eq:v_r})), we need to sum over all possible orbits passing through $R$ within the bulge, with each orbit weighted by $f_1(\bmJ,\bmTheta,t)$. To do this, we will partition the $f_1(\bmJ,\bmTheta,t)$ on a given orbit based on the value of $\theta_{\rm R}$ into two parts: one part $f_{\rm outward}$ for stars with positive radial velocities on the half orbit (pericenter to apocenter), and another part $f_{\rm inward}$ for stars with negative radial velocities on the half orbit (apocenter to pericenter). Thus, we can rewrite Equation~(\ref{eq:v_r}) as
\begin{equation}\label{eq:rhos1-vr}
    \overline{\rho_{\rm s1} v_{\rm R}} = \frac{\pi}{R^2}\int \left[ f_{\rm outward} - f_{\rm inward} \right] \d E\d L^2,
\end{equation}
and 
\begin{equation}
    \rho_{\rm s1} = \frac{\pi}{R^2}\int \frac{f_{\rm outward} + f_{\rm inward}}{|v_{\rm R}|} \d E\d L^2.
\end{equation}

Since most outflowing stars will fly out of the bulge, the star injection rate $\dot{f}_{\rm out}$ (outflow distribution) in Equation~(\ref{eq:injection_func}) does not truly correspond to the distribution function of outflow stars captured by the bulge. To address this issue, in numerical calculations of Equation~(\ref{eq:injection_func}), we set the velocity $\bm{v}$ in $\dot{f}_{\rm out}$ to be the velocity of a given orbit at position $R$
\begin{gather}\label{eq:orbital_vr_vt}
    v_{\rm R}^2 = 2\left(E-\Phi_0\right)-\frac{L^2}{R^2},\\
    v_{\perp}^2 = \frac{L^2}{R^2},
\end{gather}
where $v_{\rm R}$ and $v_{\perp}$ are the radial velocity and the non-radial velocity of star's orbital motion at $R$ respectively.
This way, the outflow distribution function actually corresponds to the probability of outflowing stars being captured by that orbit at position $R$. By integrating all orbits at a given $R$, we can obtain the corresponding density distribution of outflowing stars transforming into bulge stars
\begin{equation}
    \dot{\rho}_{\rm mix} = \int \dot{f}_{\rm out} \d v^3 = \frac{\pi}{R^2}\int \frac{\dot{f}_{\rm out}}{v_{\rm R}} \d E \d L^2.
\end{equation}
Since outflow velocity $V_{\rm out} \gg \sigma_0$, only a small fraction of outflowing stars mixes into the bulge. 
However, for slower outflows ($V_{\rm out} \sim \sigma_0$), stars formed inside outflows will mix with the bulges so that all kinematic powers of the outflows will be dissipated in the bulges. In such a case, the efficiency will be very high ($\eta\rightarrow 1$) leading to that the perturbation approximation doesn't work. We will treat this case in a future work.

\section{Numerical Results}\label{sec:results}

\begin{figure}
    \centering
    \includegraphics[width=0.98\textwidth]{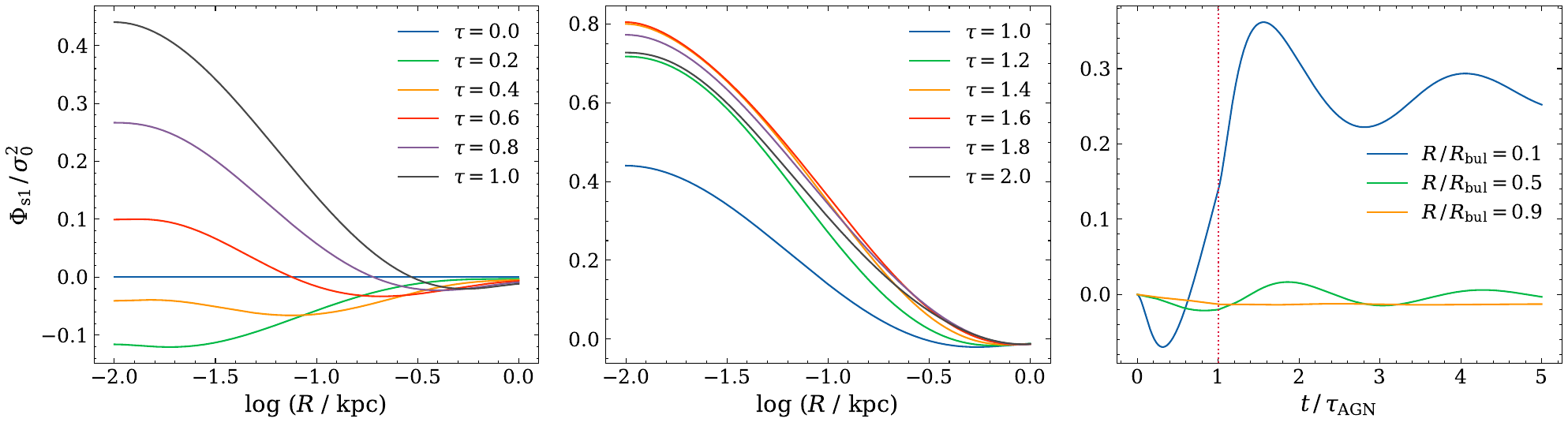}
    \caption{An example of time evolution of bulge's self-gravitational response potential $\Phi_{\rm s1}$ normalized to the $\s2$. 
    The left panel corresponds to the response within one $\tau_{\rm AGN}$, while the middle panel corresponds to the response $\Phi_{\rm s1}$ after AGN (outflow) is quenched, the right panel is the time evolution of $\Phi_{\rm s1}$ at certain radius. To show the temporal behavior, we extend the dimensionless time $\tau$ from $2$ to $5$. The vertical red dotted line indicates the time of outflow extinction.}
    \label{fig:response}
\end{figure}

For a typical galactic bulge, the velocity dispersion is $\sigma_0=\sigma_{200}\times 200\,\rm \kms$ (see Appendix~\ref{appendix:appendix.c}), we non-dimensionalize energy (and potential) by $\s2$ and velocity by $\sigma_0$. We set the velocity dispersion of outflowing stars $\sigma_{\rm e}=\sigma_0$. The outflowing velocity at $R_{\rm in}$ is $V_{\rm out,0} = V_{500}(R_{\rm in}) \times 500 \,\rm \kms$. 
We set the bulge's inner boundary $R_{\rm in}=0.01\,\rm kpc$ and outer boundary $R_{\rm bul}=1 \,\rm kpc$. 
The AGN lifetime $\tau_{\mathrm{AGN}}$ was derived by Equation~(\ref{agn_life}) for different $\dot{M}_{\rm out}$. 
The dimensionless radius and time are $r=R/R_{\rm bul}$ and $\tau = t/\tau_{\mathrm{AGN}}$, with $\tau$ ranging from $0$ to $2$. 
Other physical quantities are non-dimensionalized through the combinations of above method (such as density dimension $\s2/4\pi G R_{\rm bul}^2$).
We denote 
physical quantities with a hat without further specification. 
The bulge's total mass is $M_{\rm bul}=2\s2 R_{\rm bul}/G \approx 2\times 10^{10}\sunm\,\sigma_{200}^2R_{\rm bul,1kpc}$ (see Appendix~\ref{appendix:appendix.c}). 
%
\begin{figure}
    \centering
    \includegraphics[width=0.98\textwidth]{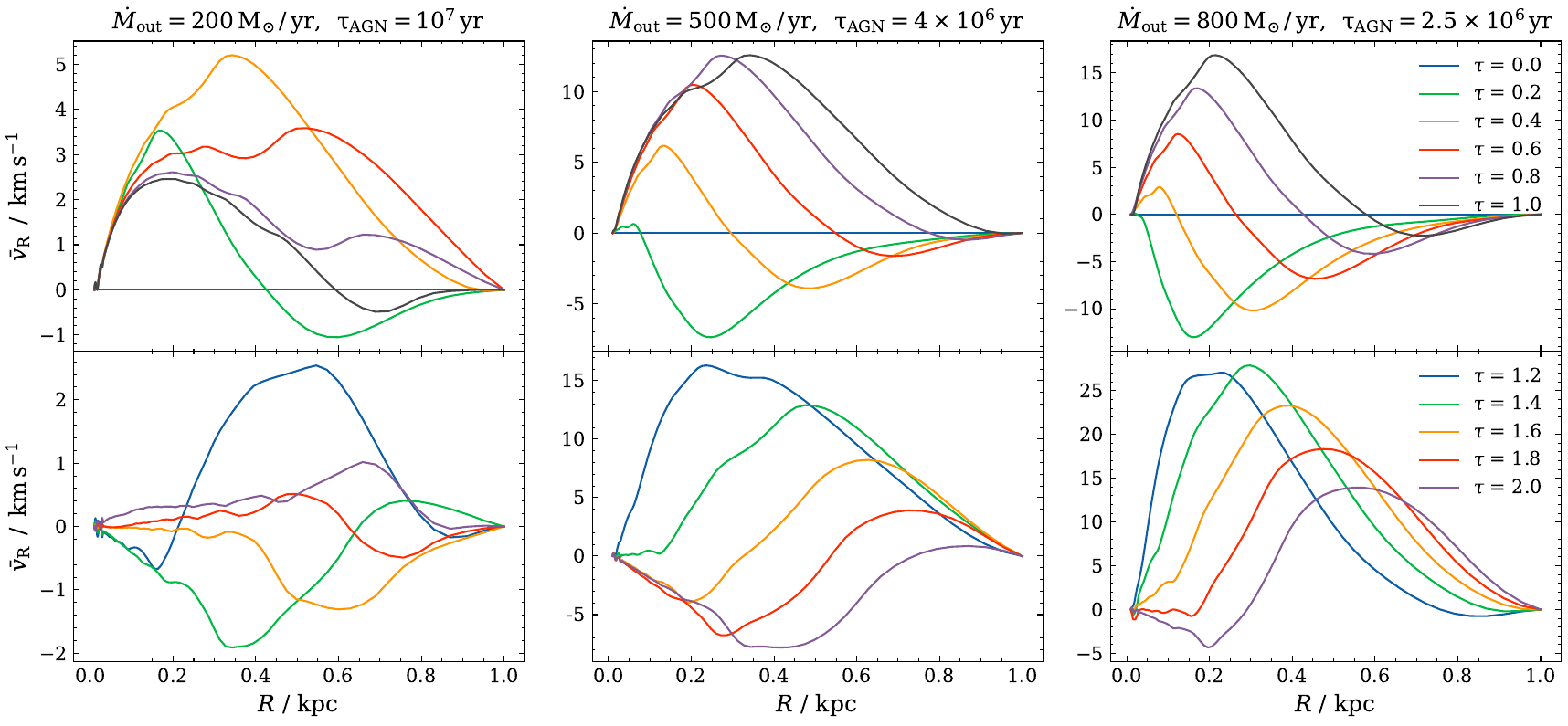}
    \caption{Radial velocity evolution for different mass injection rate and $\tau_{\rm AGN}$ with $V_{\rm out}=900\,\rm km/s$ at $R_{\rm in}$. The upper panel is the velocity evolution during $\tau_{\rm AGN}$, and the lower panel is the velocity evolution after central SMBH's active timescale. $\tau$ is the dimensionless time. We can see that with larger $\dot{M}_{\rm out}$, although the corresponding $\tau_{\rm AGN}$ is smaller, the radial velocity is still larger.
During $\tau_{\rm AGN}$, we can find that the bulge's radial velocity exhibits the similar profile for the same $\tau\times \tau_{\rm AGN}$, such as $\tau=0.4$ for $\dot{M}_{\rm out}=200\,M_\odot\,/\,\rm yr$ and $\tau=1.0$ for $\dot{M}_{\rm out}=500\,M_\odot\,/\,\rm yr$, but with different values. 
}
\label{fig:nonpowerlaw_vr_dotm1}
\end{figure}

In Figure\,\ref{fig:response}, we present the time evolution of the bulge's response potential $\Phi_{\rm s1}$ obtained by solving Equation~(\ref{eq:matrix-equation-grid}). The left panel corresponds to the evolution during the active phase of the AGN, while the middle panel corresponds to the evolution after AGN quenched. 
And the right panel is the time evolution of $\Phi_{\rm s1}$ at certain radius. To show the temporal behavior, we extend the dimensionless time $\tau$ from $2$ to $5$. 
At $\tau=0$, only $\Phi_{\rm e,out}$ in the perturbation is effective, explaining the initial negative response of $\Phi_{\rm s1}$. As the $\Phi_{\rm e,bul}$ increases linearly over $\tau$, we can observe that $\Phi_{\rm s1}$ increases, resulting in a positive response. When the outflow is quenched (vertical red dotted line in the right panel), the bulge's self gravity begins to oscillate with damping, induced by the bulk radial oscillation of bulge stars, until it finally reaches new equilibrium. 
Such effects of feedback on bulge structure and dynamics seems to be supported by the observed tight relationship between galaxy binding energy and stellar mass that likely traces feedback energy \citep{Shi2021}. 

\begin{figure}
    \centering
    \includegraphics[width=0.98\textwidth]{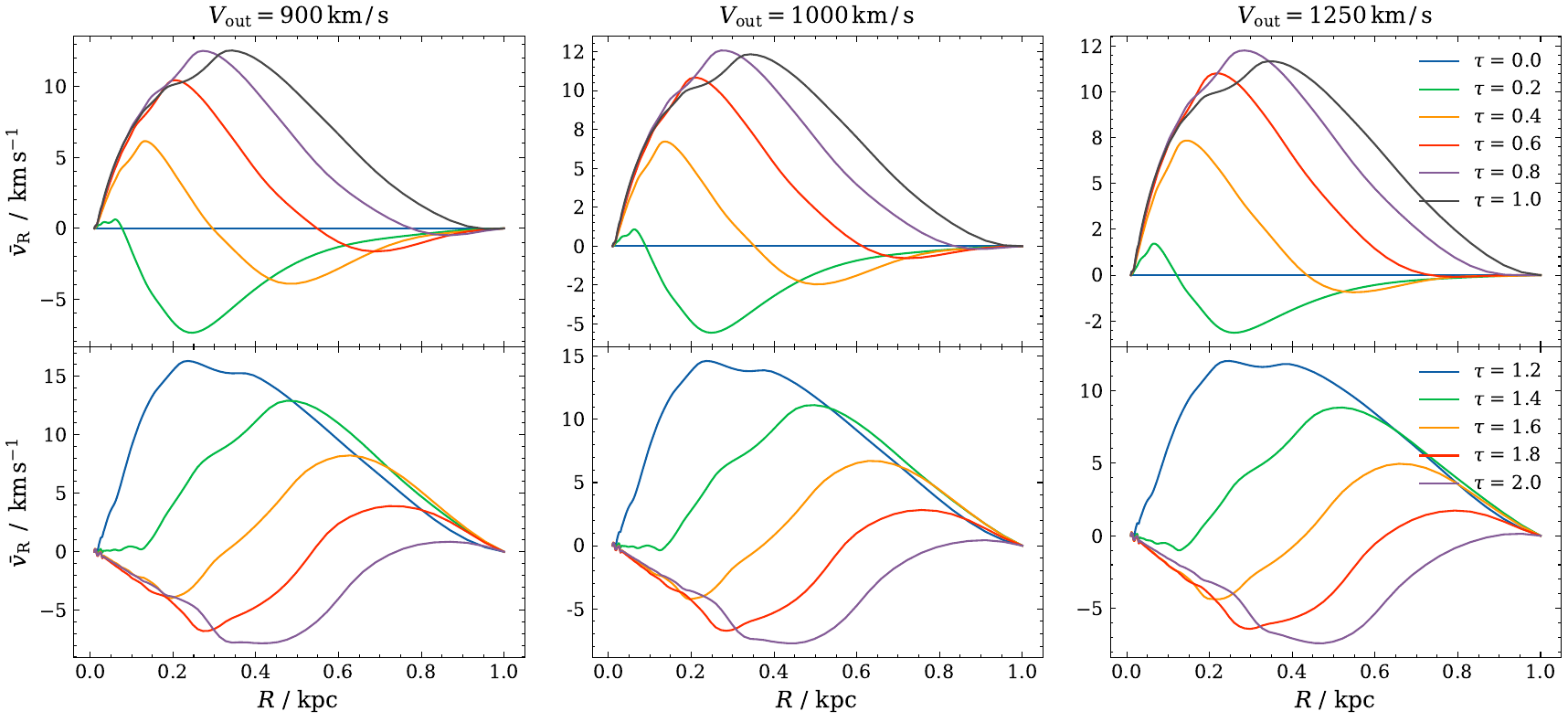}
    \caption{Radial velocity evolution for different outflowing velocity at $R_{\rm in}$ with $\dot{M}_{\rm out}=500\,\rm M_{\odot}\,/\,yr$. Same as Fig~\ref{fig:nonpowerlaw_vr_dotm1}, with each column corresponding to different outflowing velocity. With larger $V_{\rm out}$, the radial velocity is smaller. This is because with larger $V_{\rm out}$, the perturbing density is smaller (see $\rho_{\rm out,0}$), and so the response of bulge stars. And the probability of outflowing stars being captured by the bulge also decreases. 
    }
    \label{fig:nonpowerlaw_vr_V500}
\end{figure}

Using the numerical method described in Section~\ref{sec:numerical_method}, once we get $B_{\bmalpha}(t)$ from Equation\,(\ref{eq:matrix-equation-grid}), we can calculate the response density $\rho_{s1}$ momentum $\overline{\rho_{\rm s1} v_{\rm R}}$ and others, then finally the bulge's bulk velocity $\bar{v}_{\rm R}$
\begin{equation}
    \bar{v}_{\rm R} = \frac{4\pi\sigma_{0}\widehat{\overline{\rho_{\rm s1} v_{\rm R}}}}{\hat{\rho}_0 + 4\pi\hat{\rho}_{s1}},
\end{equation}
which are shown in Figure~\ref{fig:nonpowerlaw_vr_dotm1} and Figure~\ref{fig:nonpowerlaw_vr_V500}. 
Overall, similar to the case of $\Phi_{\rm s1}$, there is a brief period after $\tau=0$ during which the effect of the $\Phi_{\rm e,out}$ term causes the bulge to move inwardly. When the $\Phi_{\rm e,bul}$ term begins to take significant roles, the bulge acquires an outward velocity. The reason is apparent: as the bulge loses its gas mass through outflows beyond its scope, the bulge has to expand to reach a new equilibrium.

We can see that the larger mass injection rate, the faster the radial velocity  during both outflow on-and-off phases; while with larger outflowing velocity, the radial velocity is even smaller during outflow on and off phase; because larger $V_{\rm out}$ corresponds to smaller perturbing density (see Equation\,\ref{eq:outflow_conserv}) for a given mass injection rate $\dot{M}_{\rm out}$, the response of bulge stars is smaller. 
It can be found from the lower panels of Figure\,\ref{fig:nonpowerlaw_vr_dotm1} and Figure~\ref{fig:nonpowerlaw_vr_V500} that after AGN is quenched, with increasing in time $\tau$, value of radial velocity continuously decreases, and eventually the bulge reaches a new equilibrium, at which point the radial velocity approaches zero. 
We should note that bulge stars still possess radial velocities for a considerable period of time even after the outflow is extinguished, and the trend of radial velocity changing with mass outflow rate and outflow velocity is similar to what was described above. This may be one of the potential evidence suggesting the existence of massive, fast outflows in galaxies in the past.

\section{Conclusion}\label{sec:conclusion}

In this paper, we have calculated the radial velocity perturbation due to outflows and star formation in outflows during AGN episodes. We use the linearized perturbation theory of the Fokker-Planck equation including source and dynamical friction term to calculate the time evolution of bulge's response to outflow perturbation. We assume the perturbation and response are spherical symmetric and consider the effect of outflowing stars mixing into bulge stars by an injection rate in the Boltzmann equation. The radial velocity follows the outflow on-and-off phases, indicating a relation to the central SMBH duty cycle. We find that the radial velocity exceeds a few $10 \rm \kms$ with mass outflow rate $> 500\,\rm M_{\odot} yr^{-1}$, even after AGNs are quenched. We found that the radial velocity is not so dependent on the outflow velocity. 

Observational tests of the present model of bulge oscillations are feasible, though the model is for an ideal case with a sphere of bulges without rotation. 
First, it is possible to measure stellar velocities of the Galactic bulge as the nearest one \citep[e.g.,][]{Laurikainen2016}. 
The Milky Way was an AGN \citep[e.g.,][]{Levin2003,Levin2007,Guo2012} so that outflows developed by power radiation will lead to expansion of the Galactic bulge. 
It is thus expected to contract bulge of our own galaxy from APOGEE data of SDSS and {\textit{Gaia}} observations \citep{Queiroz2021,Zhou2021,Clarke2022} since the SMBH activity was quenched.
Second, the galactic bulge oscillations could be detected by IFU (integral-field unit) observations, such as of M\,31 \citep{BlanaDiaz2018}, NGC\, 1097, NGC\,4608 and NGC\,4643 \citep{Kolcu2023,Erwin2021}, and other nearby galaxies \citep{Rigamonti2024}. 
We may measure the shifts of emission or absorption lines of the bulges. 
By integrating the radial velocities of all stars, we can obtain the total mean velocity of the bulges. 
Radial oscillations of bulges may give rise to observational spectral consequences of line shifts according to pixels of IFU observations, from line redshifts and blue shifts, or asymmetry of redshift and blueshift components due to optical depth.

\begin{acknowledgments}
The authors greatly thank the referee for a thoughtful report improving the manuscript, in particular, on the potential perturbations. We appreciate discussions with members of IHEP AGN group, and G Zhao and Y.-J. Peng. J.M.W. thanks the support by the National Key R\&D Program of China through grants 2020YFC2201400 by NSFC-11991050, -11991054, -12333003. LCH was supported by the National Science Foundation of China (11991052, 12233001), the National Key R\&D Program of China (2022YFF0503401), and the China Manned Space Project (CMS-CSST-2021-A04, CMS-CSST-2021-A06).
\end{acknowledgments}

\appendix

\section{Angle-action Coordinate}\label{appendix:appendix.a}

Consider the 6-dimensional phase space consists of the coordinate $\bmTheta, \bmJ$, with each component \citep{Tremaine1984, Palmer1994, BT2008}
\begin{equation}
    J_{i}=\frac{1}{2\pi}\oint_{\mathcal{L}}p_{i}\d R = \frac{1}{\pi}\int_{\Rm}^{\Rp}p_{i}\mathrm{d}R,
\end{equation}
where $p_{i}$ is the $i$-component of momentum, and $\oint_{\mathcal{L}}$ indicates integration along stars' orbits with angle from $0$ to $2\pi$.
Since we only consider spherical symmetric perturbations in this article, thus in spherical coordinates $(R,\theta,\phi)$, the three independent action components $J_{i}$ are $(J_{R},J_{\theta},J_{\phi})$, which can be chosen as $(J_{R},L,L_{z})$ \citep[see Table 3.1 in][]{BT2008}, with $L,L_{z}$ are the angular momentum of star's orbital movement and the $z$ component of the angular momentum. We use the Jacobi function $S(\bm{x},\bmJ)$ to get the conjugate angle coordinates, where $(\bm{x},\bm{v})$ and $(\bmTheta,\bmJ)$ are conjugate coordinates, respectively
\begin{align}
    S(\bm{x},\bmJ) &=\int_{q}p(q,P)\d q=\int p_{R}dR +\int p_{\theta}\d \theta + \int p_{\phi}\d \phi \\ &= \int_{\mathcal{L}}\d R\sqrt{2(E-\Phi)-\frac{L^2}{R^2}} + \int_{\mathcal{L}}\d \theta \sqrt{L^2-\frac{L_z^2}{\sin^2\theta}}+L_{z}\phi,
\end{align}
the components of momentum in spherical coordinates are
\begin{align}
    & p_{\rm R} = \dot{R}, \\
    & p_{\theta} = R^2\dot{\theta}, \\
    & p_{\phi} = R^2 \sin^2\theta \dot{\phi},
\end{align}
and the Hamiltonian can be written as
\begin{equation}
    H = \frac{1}{2} \left( p_{\rm R}^2+\frac{p_{\theta}^2}{R^2}+\frac{p_{\phi}^2}{R^2 \sin^2\theta} \right) + \Phi(R),
\end{equation}

Since we mainly use $(\theta_{\rm R}, J_{\rm R})$ in our calculation, we have $J_{\rm R}$, $\theta_{\rm R}$ and its relation with $E, L$
\begin{equation}
    J_{\rm R} = \frac{1}{2\pi}\oint p_{\rm R}\d R = \frac{1}{\pi}\int_{\Rm}^{\Rp}\sqrt{2(E-\Phi)-L^2/R^2} \d R,
\end{equation}
and
\begin{align}
    \frac{\partial J_{\rm R}}{\partial E} = \frac{2}{\pi}\int_{\Rm}^{\Rp}\frac{\d R}{\sqrt{2(E-\Phi)-L^2/R^2}} = \frac{1}{\Omega_{\rm R}},
\end{align}
\begin{equation}
    \frac{\partial J_{\rm R}}{\partial L} = \frac{1}{\pi}\int_{\Rm}^{\Rp}\d R\frac{L^2}{R^2\sqrt{2(E-\Phi)-L^2/R^2}} = \frac{\Omega_{\nu}}{\Omega_{\rm R}},
\end{equation}
where $\Omega_{\rm R}$ is the frequency of star's radial oscillation, $\Omega_{\nu}$ is the frequency of star's angular oscillation. If we consider $E$ as a function of $J_{\rm R}$ and $L$,
\begin{align}
    & \frac{\partial E}{\partial J_{\rm R}} = \Omega_{\rm R}, \\
    & \frac{\partial E}{\partial L} = \frac{\partial E}{\partial J_{R}}\frac{\partial J_{\rm R}}{\partial L} = \Omega_{\nu},
\end{align}
and from Jacobi function \citep{Palmer1994}
\begin{equation}
    \bmTheta = \frac{\partial S}{\partial \bmJ},
\end{equation}
so we can calculate $\theta_{\rm R}$
\begin{equation}
    \theta_{\rm R} = \frac{\partial S}{\partial J_{\rm R}} = \int_{\mathcal{L}} \frac{\Omega_{\rm R}\d R}{\sqrt{2(E-\Phi) - L^2/R^2}},
\end{equation}
Our aim is to use angle-momentum variables to simplify the Boltzmann equation, get the response matrix, then transform into real $(R,\theta,\phi)$ space.

\section{Dynamical Friction and Collision}\label{appendix:appendix.b}

The evolution of unperturbed bulge distribution is governed by the collision-less Boltzmann equation
\begin{equation}
    \frac{\partial f}{\partial t} + \bm{v}\cdot\frac{\partial f}{\partial \bm x} - \nabla\Phi\cdot\frac{\partial f}{\partial \bm v} \sim 0,
\end{equation}
where $\Phi$ is the smoothed-out potential indicating there's no collision/encounter between stars. For a typical unperturbed bulge, the two-body relaxation timescale is \citep{BT2008}
\begin{equation}
    \tau_{\rm relax} \sim \frac{0.1N_\star}{\ln{N_\star}}\frac{R_{\rm bul}}{\sigma_0} \sim 10^3\,\rm Gyr,
\end{equation}
where $N_\star\sim 10^9-10^{10}$ is the total number of stars in the bulge. Star encounter only takes significant effect on timescale exceeding bulge's lifetime $\sim 10 \,\rm Gyr$ . For weak encounter between stars, the star's original orbit is perturbed by the outflowing stellar encounters.  If we only consider local encounters, the encounter operator can be expressed under Fokker-Planck approximation as
\begin{equation}
    \Gamma(f) = - \sum_{i=1}^{3}\frac{\partial}{\partial v_{i}}\left[D(\Delta v_{i})f \right] + \frac{1}{2} \sum_{i,j=1}^{3}\frac{\partial^2}{\partial v_{i}\partial v_{j}}\left[D(\Delta v_{i}\Delta v_{j})f \right],
\end{equation}
where $v_i$ are velocity components, $D(\Delta v_{i}), D(\Delta v_{i}\Delta v_{j})$ are the diffusion coefficients, with $D(\Delta v_{i})$ describing the drifting process and $D(\Delta v_{i}\Delta v_{j})$ describing the diffusion process
\begin{gather}
    D(\Delta v_{i}) = 4\pi\ln{\Lambda}G^2m_{*}(m_{*}+M_{*})\frac{\partial}{\partial v_{i}} \int \frac{f \d^3\bm v}{|\bm V_{\rm out}-\bm v|} = D_{0}(\Delta v_{i}) + \epsilon D_1(\Delta v_{i}), \\
    D(\Delta v_{i}\Delta v_{j}) = 4\pi\ln{\Lambda}G^2m_{*}^2\frac{\partial^2}{\partial v_{i}\partial v_{j}} \int \d^3\bm v f |\bm V_{\rm out}-\bm v| = D_{0}(\Delta v_{i}\Delta v_{j}) + \epsilon D_1(\Delta v_{i}\Delta v_{j}).
\end{gather}
where $m_{*}$ is the mass of a single star in bulge, and subscript 0 or 1 describes calculation using unperturbed or perturbed distribution function, marked by $\epsilon\ll 1$. We do not consider mass distribution here, thus we take all star mass equals to $m_{*}$.

When bulge is perturbed by star formation in AGN's outflow in our case, using linear perturbation theory we can write perturbations as
\begin{equation}
    X = X_0 + X_1,
\end{equation}
where $X$ denote physical quantities, unperturbed and perturbed quantities are represented by subscripts $0$ and $1$. Under the assumption that $f_0$ is Maxwellian and outflow perturbation is spherical symmetric in Section.~\ref{sec:theory}, the non-zeros diffusion coefficients can be written as \citep{BT2008}
\begin{gather}\label{eq:D0}
    D_0(\Delta v_{\rm R}) = - \frac{4\pi\ln{\Lambda}G^2\rho_0(m_{*}+M)}{\s2}G_x, \\
    D_0\left[(\Delta v_{\rm R})^2\right] = \frac{4\sqrt{2}\pi\ln{\Lambda}G^2\rho_0 m_{*}}{\sigma_0} \frac{G_x}{x}, \\
    D_0\left[(\Delta v_{\perp})^2\right] = \frac{4\sqrt{2}\pi\ln{\Lambda}G^2\rho_0 m_{*}}{\sigma_0} \frac{{\rm erf}(X) - G_x}{x},
\end{gather}
with
\begin{equation}
    G_x = \frac{1}{2x^2}\left[{\rm erf}(x) - \frac{2x}{\sqrt{\pi}}e^{-x^2} \right].
\end{equation}
where $x=V_{\rm out}/(\sqrt{2}\sigma_0)$ and $\rm erf(x)$ is the error function, $v_{\perp}$ is the velocity perpendicular to $v_{\rm R}$. 


\section{Bulge Distribution}\label{appendix:appendix.c}

We shall note that the isothermal distribution we choose in Section.~\ref{sec:theory} only appeals to radius from $R=0$ to $R=\infty$. If we assume that the interior of bulge's inner radius $R_{\rm in}$ is hollow, then we can solve for the distribution of density and potential through Poisson's equation as
\begin{gather}
    \rho_0 = \frac{\s2}{2\pi G R^2}, \\
    4\pi G\rho_0 = \nabla^2 \Phi_0 = \frac{1}{R^2}\frac{\rm d}{\mathrm{d}R}\left(R^2\frac{\mathrm{d}\Phi_0}{\mathrm{d}R}\right).
\end{gather}
and considering the inner and outer boundary conditions as
\begin{gather}
    \frac{\mathrm{d}\Phi_0}{\mathrm{d}R}|_{R=R_{\rm in}} = 0,\\
    \Phi_0(R_{\rm bul}) = - \frac{GM_{\rm bul}}{R_{\rm bul}}.
\end{gather}
where $M_{\rm bul}$ is the bulge total mass. We can solve the above BVP (boundary value problem) and get the potential
\begin{equation}\label{eq:Phi_bvp}
    \Phi_0 = 2\s2\left[\ln{\left(\frac{R}{R_{\rm bul}}\right)} + \frac{R_{\rm in}}{R} -1 \right].
\end{equation}
when $R_{\rm in} < R < R_{\rm bul}$, which differs from the distribution we used in the previous section. 

However, if we consider the gravitational potential of the central SMBH, the above problem can be fixed. For a typical central SMBH mass $M_{\bullet} \sim 10^8 M_{\odot}$, the bulge's total mass approximates to $M_{\rm bul} \sim 2\times 10^{10} M_{\odot}$ \citep{Kormendy2013}, meanwhile using the bulge's total mass from our model
\begin{equation}
    M_{\rm bul} = \int^{R_{\rm bul}}_{R_{\rm in}} 4\pi R^2 \rho_0 \mathrm{d}R = \frac{2\sigma_0^2}{G}\left(R_{\rm bul} - R_{\rm in} \right)
    \approx 2\times 10^{10}\sunm\,\sigma_{200}^2R_{\rm bul,1kpc}.
\end{equation}
gives the $\sigma_0 \sim 200 \,\rm \kms$, where $R_{\rm bul,1kpc}=R_{\rm bul}/1\,\rm kpc$. And using the $M_{\bullet}-\sigma$ relation \citep{Gebhardt2000, Ferrarese2000, Kormendy2013}, for a SMBH with a mass of $M_{\bullet}\sim 10^8M_{\odot}$, the approximate velocity dispersion $\sigma\sim 200\,\rm \kms$, which is coherent in former approximation, and verified that our assumption about the distribution is self-consistent. We choose $R_{\rm in} = 0.01 R_{\rm bul}$, and we can rewrite the second term on the RHS of Equation~(\ref{eq:Phi_bvp}) as
\begin{equation}
    -\frac{2\s2 R_{\rm in}}{R} \simeq -\frac{GM_{\bullet}}{R} ,
\end{equation}
thus once we consider both the bulge potential and the central SMBH's potential, the initial distribution assumption in Section.~\ref{sec:theory} can be justified.

\bibliography{sample631}{}
\bibliographystyle{aasjournal}

\end{document}